\documentclass[twocolumn,prl,showpacs,preprintnumbers,amsmath,amssymb]{revtex4-1}

\usepackage{graphicx}% Include figure files
\usepackage{dcolumn}% Align table columns on decimal point
\usepackage{bm}% bold math
\usepackage{subfigure}

\begin{document}

\preprint{APS/123-QED}

\title{Preferential antiferromagnetic coupling of vacancies in graphene on SiO$_2$:\\ Electron spin resonance and scanning tunneling spectroscopy}
\author{ S. Just$^{1}$, S. Zimmermann$^2$,
%M. Grob$^1$, N. Freitag$^1$, K. Fl\"ohr$^1$,
  V. Kataev$^2$, B. B\"uchner$^2$, M. Pratzer$^1$ and M. Morgenstern$^1$}
\affiliation{ $^{1}$II. Institute of Physics B and JARA-FIT, RWTH Aachen, 52074 Aachen, Germany\\ $^{2}$ IFW Dresden, Institute of Solid State and Materials Research, D-01069 Dresden, Germany}

\date{\today}

\begin{abstract}
Monolayer graphene grown by chemical vapor deposition and transferred to SiO$_2$ is used to introduce vacancies by Ar$^+$ ion bombardment at a kinetic energy of 50 eV.
The density of defects visible in scanning tunneling microscopy (STM) is considerably lower than the ion fluence implying that most of the defects
are single vacancies as expected from the low ion energy.
The vacancies are characterized by scanning tunneling spectroscopy (STS) on graphene and HOPG. A peak close to the Dirac point is found within the local density of states of the vacancies similar the the peak found previously for vacancies on HOPG. The peak persists after air exposure up to 180 min, such that electron spin resonance (ESR) at 9.6 GHz can probe the vacancies exhibiting such a peak. After an ion flux of 10/nm$^2$, we find an ESR signal corresponding to a $g$-factor of 2.001-2.003 and a spin density of  1-2 spins/nm$^2$. The peak width is as small as 0.17 mT indicating exchange narrowing. Consistently, the temperature dependent measurements reveal antiferromagnetic correlations with a Curie-Weiss temperature of -10 K. Thus, the vacancies preferentially couple antiferromagnetically ruling out a ferromagnetic graphene monolayer at ion induced spin densities of $1-2$/nm$^2$.\end{abstract}

\maketitle
\section{Introduction}
Defect induced magnetism is controversially discussed based on indications for ferromagnetism in oxides, nitrides, sulfides, and carbon based materials \cite{Ogale,Esquinazi1,Popa}. Graphite or graphene might be the most simple candidate of them, since it contains only one element, is structurally simple, and is rather inert.
The theoretical prediction of interacting magnetic moments provided by vacancies \cite{Lehtinen,Carlson,Yazyev,Lopez,Pisani,Wessel,Li} and zig-zag edges \cite{Fujita, Myamoto,Yu,Son} fuels the hope that magnetic order can be achieved.
However, experimental evidence for paramagnetism \cite{Grigorieva1,Grigorieva2,Ricco,Lee}, ferromagnetism \cite{Esquinazi1, Esquinazi2, Cervenka, Wang, Mombru, Rao3, Esquinazi2b, Esquinazi2c, Esquinazi3, Esquinazi4,Esquinazi5} and antiferromagnetism \cite{Shibayama,Kaustelkis,Yablokov1,Yablokov2,Joly} in graphene and graphite appear to contradict each other, even though partly found by different experimental methods after different sample preparation. For example, early Superconducting Quantum Interference Device (SQUID) measurements on Highly Oriented Pyrolytic Graphite (HOPG) found indications for ferro- or ferrimagnetism after irradiation with 2.25 MeV protons even at room temperature \cite{Esquinazi2}. Subsequent, more detailed investigations observed ferromagnetism also for N$^{4+}$- and C$^{4+}$-projectiles and indirectly concluded that the ferromagnetism requires a particular vacancy-vacancy distance of about 2 nm \cite{Esquinazi3}. The conclusion is based on SRIM calculations \cite{TRIM} and the observation of some x-ray magnetic circular dichroism (XMCD) features observed at 300 K, which imply near-surface magnetism \cite{Esquinazi5}.
In contrast, a recent report on SQUID results studying graphene laminates after bombardment with 10$^{20}$/m$^2$ protons of kinetic energy of 350-400 keV or $5\cdot 10^{17}-10^{20}$/m$^2$ C$^{4+}$ ions with kinetic energy of 20 MeV finds only paramagnetic spin 1/2 centers down to 1.8 K \cite{Grigorieva1}, albeit the ion fluence and ion energy of C$^{4+}$ matches rather exactly the ones leading to ferromagnetism in HOPG \cite{Esquinazi3}.  It was further found, that
the spin 1/2 centers come in two types distinct by their doping behaviour, which are probably caused by unsaturated $\pi$-type and $\sigma$-type electrons, respectively \cite{Grigorieva2}.
Also thicker graphene samples (2 nm) vertically stacked on a Si substrate did not show any magnetic hysteresis
after 100 keV N$^+$ bombardment up to ion densities of 10$^{21}$/m$^2$ and down to temperatures of 5 K \cite{Ney}.
One drawback of these studies is that the damage caused by the ions is only estimated by SRIM simulations \cite{TRIM},
which ignore the crystalline structure of the honeycomb lattice and any type of annealing either caused by temperature or by subsequent ions. For example, the number of paramagnetic centers deduced by SQUID was only about 10 \% of the calculated ion induced vacancies \cite{Grigorieva1}. On the other hand, a clear fingerprint of paramagnetic vacancies in scanning tunneling microscopy (STM) has been reported, which is a peak in the local density of states (LDOS) close to the Dirac point $E_D$ \cite{Brihuega}, which according to tight binding calculations persists up to vacancy densities of 5 \% \cite{Pereira}. A peak at $E_D$ has also been found after dilute H adsorption, e.g. in density functional theory calculations \cite{Nieminnen} or in STM experiments \cite{Hess}, again indicating paramagnetic behavior.\\
Here, we combine STM and electron spin resonance (ESR) measurements on the same monolayer graphene samples after low energy ion bombardment.
%This provides direct comparison between the local electronic structure of the induced defects and their magnetic response.
ESR has the advantage with respect to SQUID that it can distinguish between different magnetic impurities by their $g$-factor and their hyperfine interaction. Thus, ESR is much less prone to unwanted ferromagnetic inclusions than SQUID. ESR has been applied previously to graphene samples as  exfoliated graphene including monolayers on scotch tape \cite{Ciric}, reduced graphene oxide \cite{Ciric2,Ciric3,Su}, or gas phase produced graphene platelets \cite{Yablokov1,Yablokov2}. Defects were present in all samples as deduced from the ESR signals, but their origin and type remains unknown. ESR has also been applied to different nanographitic structures with rather uncontrolled thickness distributions, where signals are interpreted in terms of vacancies, edge states and itinerant electrons \cite{Kaustelkis, Smith, Shibayama, Lijewski, Andersson, Kempinski, Joly, Joly2, Rao1, Rao2, Rao3, Rao4, Tommasini, Fabian, Ciric4, Akbari}, but again the origin and type of the defects was unknown. Recently the first electrically detected ESR study on heavily doped graphene on SiC has been published \cite{deHeer} revealing a contrast in conductivity $\Delta \sigma/\sigma$ of about 0.5 \% for the conduction electrons, which was used to pinpoint the strength of valley splitting in graphene on SiC. \\
Thus, neither an ESR study of graphene after controlled introduction of defects nor a study combining scanning tunneling spectroscopy (STS) and ESR has been published previously. Here, we provide such a study.\\
Firstly, we argue that the ion bombardment produces single vacancies within the graphene. The arguments are as follows:
\begin{itemize}
\item
The ions do not have enough energy to produce a second vacancy, if the displacement energy calculated by density functional theory (DFT) is correct \cite{Krashen}.
\item
The defect yield observed by STM is $Y=0.1$ as expected from SRIM, thus much lower than one implying that only a fracture of the ions displaces C atoms.
This makes it very unlikely that two C atoms are displaced in one ion impact event.
\item
Vacancies in graphene do not move at room temperature according to a recent transmission electron microscopy (TEM) study \cite{Kaiser} in accordance with results from DFT \cite{Krashen}.
\item
The defects show a LDOS peak close to the Dirac point $E_{\rm D}$, which is expected for single vacancies \cite{Pereira,Brihuega}, but not for vacancy agglomerates \cite{Lherbier,Ugeda}.
\end{itemize}
Secondly, we find that the defect-related LDOS peak close to $E_{\rm D}$, which is consistent with a paramagnetic behavior of the vacancy \cite{Lehtinen,Pereira,Brihuega}, survives exposure to air for about 3 hours, such that the transfer to the ESR setup does not destroy this characteristic spectroscopic feature. Consequently, LDOS peaks observed close to $E_{\rm {\rm D}}$ can be correlated to the ESR signal.
Thirdly, we perform ESR on the graphene with vacancies, which exhibits a peak corresponding to a $g$-factor of about 2.002. A small anisotropy of the resulting $g$-factor between in-plane and out-of-plane magnetic fields (0.02\%) has been found, which might be helpful to identify the $\pi$ or $\sigma$ character of the paramagnetic spins.
The narrow ESR linewidth of 0.2 mT at a spin density of $1-2$/nm$^2$ indicates a significant exchange narrowing and, indeed, by temperature dependent
ESR measurements, we find that the vacancies are correlated antiferromagnetically, i.e. the Curie-Weiss temperature is about -10 K. This rules out ferromagnetism in graphene at ion induced spin densities of $1-2$/nm$^2$.\\
\begin{figure*} [t!]
\includegraphics[width=17.8cm]{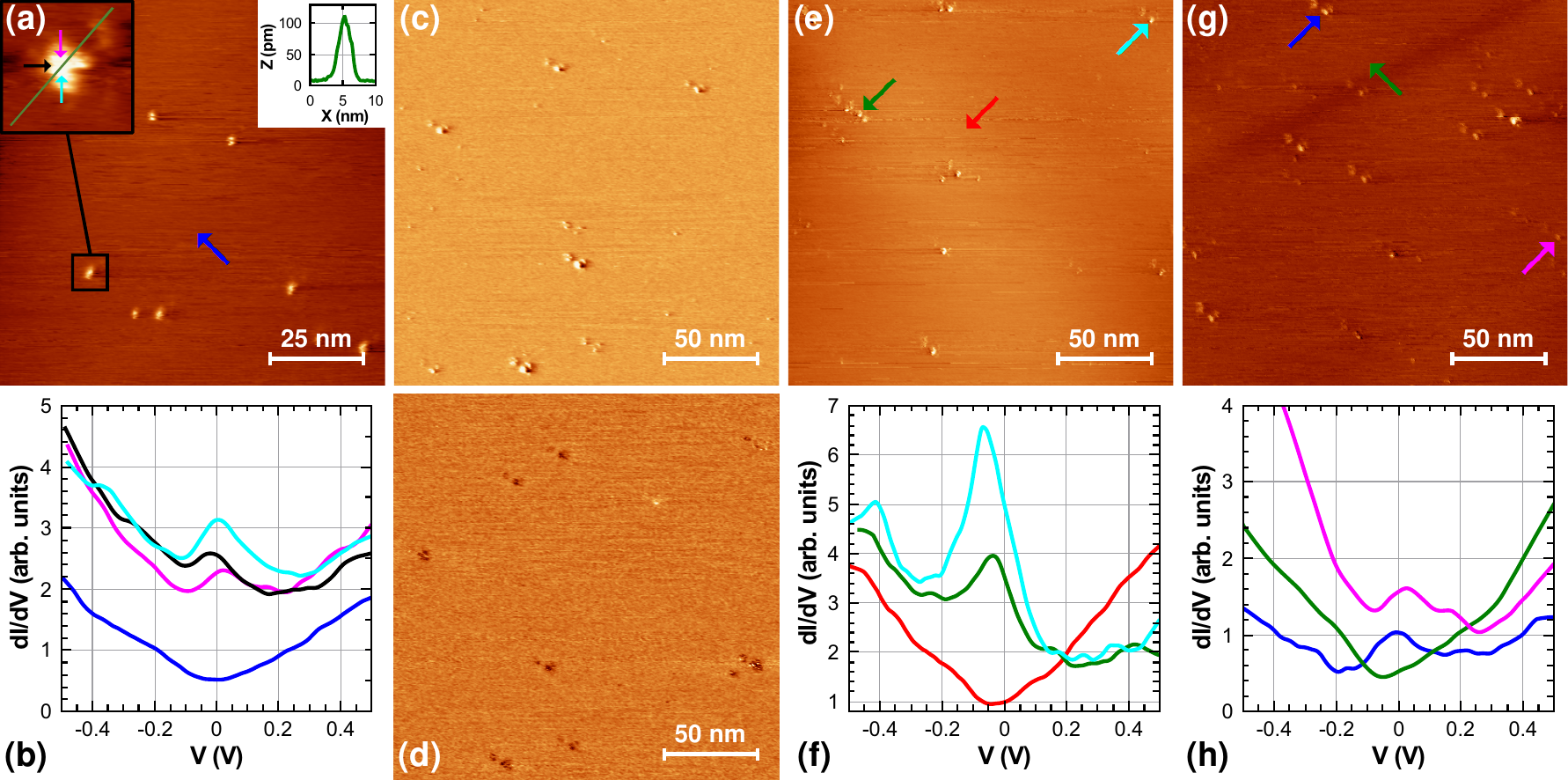}
\caption{(color online) {\it Ar$^+$ bombardment on HOPG,  $E_{\rm kin}=50$ eV, fluence: $7.4\cdot 10^{-3}$ ions/nm$^2$}: (a) STM image with arrow marking the position where the dark blue $dI/dV$ curve in (b) is recorded, $I=0.5\rm\,nA$, $V=700\rm\,mV$ ; left inset: STM image of single defect with arrows marking the positions where the $dI/dV$ curves displayed with identical color in (b) are recorded and green line marking the profile line displayed in the right inset, $I=1 \rm\,nA$, $V=700\rm\,mV$; right inset: profile line across a single defect along the line marked in the left inset; (b) $dI/dV$ curves recorded on the positions marked in (a), pink, black and light blue curve: $I_{\rm stab}= 1$ nA, $V_{\rm stab}=700$ mV, $V_{\rm mod}=8$ mV, dark blue curve: $I_{\rm stab}= 0.5$ nA, $V_{\rm stab}=700$ mV, $V_{\rm mod}=8$ mV; (c) $dI/dV$ image recorded within a different area than (a), $I=0.015\rm\,nA$,  $V=-10\rm\,mV$; (d) $dI/dV$ image of the same area as in (c), $I=0.08\rm\,nA$, $V=-160\rm\,mV$; (e) STM image after 10 min of air exposure with arrows marking the positions of $dI/dV$ curves in (f), $I=0.7\rm\,nA$, $V=600\rm\,mV$; (f) $dI/dV$ curves recorded at the positions marked in (e), red curve: $I_{\rm stab}= 0.7$ nA, $V_{\rm stab}=600$ mV, $V_{\rm mod}=10$ mV, green curve: $I_{\rm stab}= 0.8$ nA, $V_{\rm stab}=600$ mV, $V_{\rm mod}=10$ mV, light blue curve: $I_{\rm stab}= 1$ nA, $V_{\rm stab}=700$ mV, $V_{\rm mod}=10$ mV; (g) STM image after 3 hours of air exposure with arrows marking the positions of $dI/dV$ curves in (h), $I=0.2\rm\,nA$, $V=500\rm\,mV$; (h) $dI/dV$ curves recorded at the positions marked in (g), blue and green curve: $I_{\rm stab}= 0.2$ nA, $V_{\rm stab}=500$ mV, $V_{\rm mod}=10$ mV, pink curve: $I_{\rm stab}= 0.1$ nA, $V_{\rm stab}=-100$ mV, $V_{\rm mod}=10$ mV.}
\label{Fig1}
\end{figure*}
\section{Experiment}
The STM/STS experiments were performed in Ultra High Vacuum (UHV) at a base pressure of 10$^{-8}$ Pa. The graphene samples are grown by chemical vapor deposition on a Cu foil and transferred by a wet PMMA based process to a Si substrate covered with 290 nm of SiO$_2$ and Au/Ti contacts for electrical measurements \cite{2DTech}. The samples are prepared ex-situ and the monolayer thickness has been checked by Raman spectroscopy revealing an intensity ratio between the D peak and the G peak below 0.1 and a peak width of the 2D peak of 25 cm$^{-1}$. The four Au contacts are wire bonded revealing a sheet resistivity of the graphene of 3 k$\Omega$. HOPG samples used for comparative experiments are cleaved in-situ. Both type of samples were bombarded by Ar$^+$ ions of 50 eV produced by an ion plasma gun \cite{Liebmann}. The Ar pressure at the sample during ion bombardment was $5\times 10^{-3}$ Pa. Within five minutes after bombardment, the sample is transferred to another UHV chamber separated from the preparation chamber by a UHV valve and exhibiting $p=10^{-8}$ Pa. The ion flux is calibrated by measuring the current on a steel plate leading to a good estimate, since secondary electron emission has a rate of about 0.01 electrons/ion at 50 eV only and even possible O$^-$ sputtering has a rate of only 0.1/ion \cite{Walton}.\\
The STM measurements are performed with a modified Omicron STM operating at room temperature with the voltage $V$ applied to the sample. Spectroscopic $dI/dV$ curves and images are obtained by lock-in technique using an additional modulation voltage $V_{\rm mod}$. For $dI/dV$ curves, the tip is stabilized at voltage $V_{\rm stab}$ and current $I_{\rm stab}$ prior to opening the feedback loop.\\
The samples are transferred to the ESR setup within a sealed glass tube which has been cleaned by rinsing in HCl, deionized water, and acetone as well as an Ar plasma discharge. Afterwards, the samples are taken out of UHV, are mounted to the quartz glass based ESR sample holders, contacted by wire bonding and put into the glass tube, which is evacuated to $5\cdot 10^{-4}$ Pa before being filled with Ar gas. At the ESR setup the samples are removed from the tube and are directly mounted including necessary contacts. The setup is shortly pumped to $2\cdot10^4$ Pa prior to cooling to a temperature $T=4$ K, which realizes a cryovacuum. Altogether, the sample is exposed to ambient conditions including the time until the ESR setup is cooled to 4 K for below one hour.\\
The ESR measurements are performed with a standard X-band spectrometer from Bruker operating at a frequency $f \simeq 9.6$ GHz in magnetic fields up to $B=0.9$ T. The $B$ field is additionally modulated by a small ac field with the amplitude $B_{\rm mod}$ at frequency $f_{\rm mod} = 100$ kHz. This enables the use of lock-in technique, such that the detected signal shows the derivative of the reflected microwave power $dP/dB$. The sample is placed in the middle of a rectangular shaped resonator working in the TE102 mode. For the quantitative determination of the number of spins, a reference crystal made of ruby was used as described in \cite{Chang} and also shortly below. The spectrometer is equipped with a goniometer for rotating the sample with respect to the external magnetic field. A continuous-flow liquid-helium cryostat offers variable temperatures down to $T=3.7$ K.\\
For the realization of electrically detected spin resonance (EDSR), we use a battery current source and a preamplifier, with the latter connected to the lock-in amplifier. These components are especially suited for low noise applications, such that relative current changes down to $\Delta I/I_0=10^{-5}$ can be detected \cite{Lipps}.\\

\section{Results and Discussion}
\subsection{STM results on HOPG}

\begin{figure} [t!]
\includegraphics[width=7.0cm]{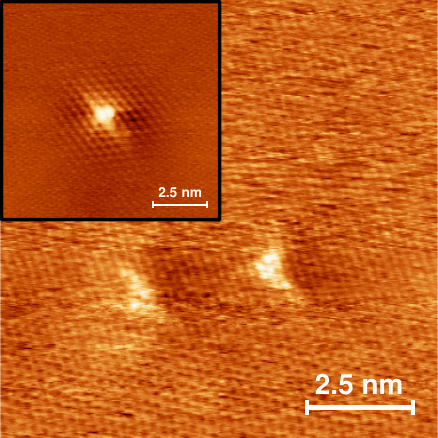}
\caption{(color online) Atomically resolved STM image of two defects after Ar$^+$ bombardment at  $E_{\rm kin}=50$ eV, fluence: $3\cdot 10^{-3}$ ions/nm$^2$,  $I=0.05\rm\,nA$, $V=-50\rm\,mV$; inset: higher resolution image of a single defect exhibiting the $\sqrt{3}\times\sqrt{3}$ superstructure around the defect more clearly, $I=0.2 \rm\,nA$, $V=700\rm\,mV$.}
\label{Fig0}
\end{figure}

We firstly performed experiments on HOPG profiting from the flat sample surface. Fig. \ref{Fig1}(a) shows a STM image after an ion fluence of $7.4 \cdot 10^{-3}$/nm$^2$. Defects are discernable as white bumps which do not appear prior to ion bombardment. A height profile of such a bump is given in the right inset exhibiting an apparent height of about 100 pm. The apparent height fluctuates between 70 pm and 150 pm from defect to defect. The averaged diameter of the bumps is 1.8 nm. Similar protrusions interpreted as single vacancies have been found previously after ion bombardment or H treatment of HOPG \cite{Coratger,Brihuega,Kang} and are predicted theoretically for single vacancies \cite{Mizes}. Protrusions of this size have also been calculated for more complex defect structures involving several vacancies \cite{Cockayne} as found, e.g. for graphene on SiC \cite{Stroscio}. On SiC, larger defects mostly involving the lower interface of the graphene can also appear as depressions \cite{Mallet}.
Typically, the elevations in STM images that belong to single vacancies exhibit a triangular appearance with an extension of about 2 nm surrounded by a $\sqrt{3}\times\sqrt{3}$ superstructure \cite{Coratger,Brihuega,Mizes}. Fig. \ref{Fig0} shows a higher resolution image of HOPG after low-energy Ar$^+$ bombardment confirming such an appearance with a triangular central structure surrounded by a $\sqrt{3}\times\sqrt{3}$ superstructure.\\
The averaged defect density obtained from several images as Fig. \ref{Fig1}(a) is $8 \cdot 10^{-4}$/nm$^2$. This is a factor of 10 smaller than the ion fluence indicating a low yield of $Y\simeq 0.1$, i.e. the onset of defect formation at 50 eV. A similar defect yield after Ar$^+$ bombardment at 50 eV has been observed previously by STM on HOPG \cite{Kang} and by molecular dynamics (MD) simulation on graphene \cite{Lehti}. The MD data did not show any double vacancy formation by the ion impact. Notice that the displacement yield on amorphous carbon as determined by SRIM \cite{TRIM} is only 0.15 C atoms/ion implying that nearly every displaced atom remains as a vacancy. These two results imply that the low energy ion impacts produce, at most, a single vacancy. In line, DFT predicts a formation energy of a relaxed vacancy within graphene of $E_{\rm form}=7.4$ eV  \cite{Krashen}, such that a single Ar$^+$ ion has not enough energy to produce a second vacancy due to the very different masses of Ar and C and the correspondingly low kinematic factor of 0.15. A migration of the vacancies into vacancy clusters is also unlikely, since the migration barrier for vacancies as deduced from DFT is $E_{\rm Dif}=1.3-1.7$ eV \cite{Krashen} implying that the vacancies are immobile at room temperature, e.g. yielding a hopping rate of $\nu_0\times \exp{(-E_{\rm Dif}/k_{\rm B}T)}\simeq 10^{-(10-17)}$/s assuming a reasonable attempt frequency $\nu_0=10^{13}$/s ($k_{\rm B}$: Boltzmann constant). In line, a recent transmission electron microscopy (TEM) analysis reveals that single vacancies produced by H$^+$ bombardment do not move at room temperature after a Jahn-Teller type reconstruction, if protected from radiation damage by the electron beam \cite{Kaiser}.\\
\begin{figure*} [htb]
\includegraphics[width= 13.35cm]{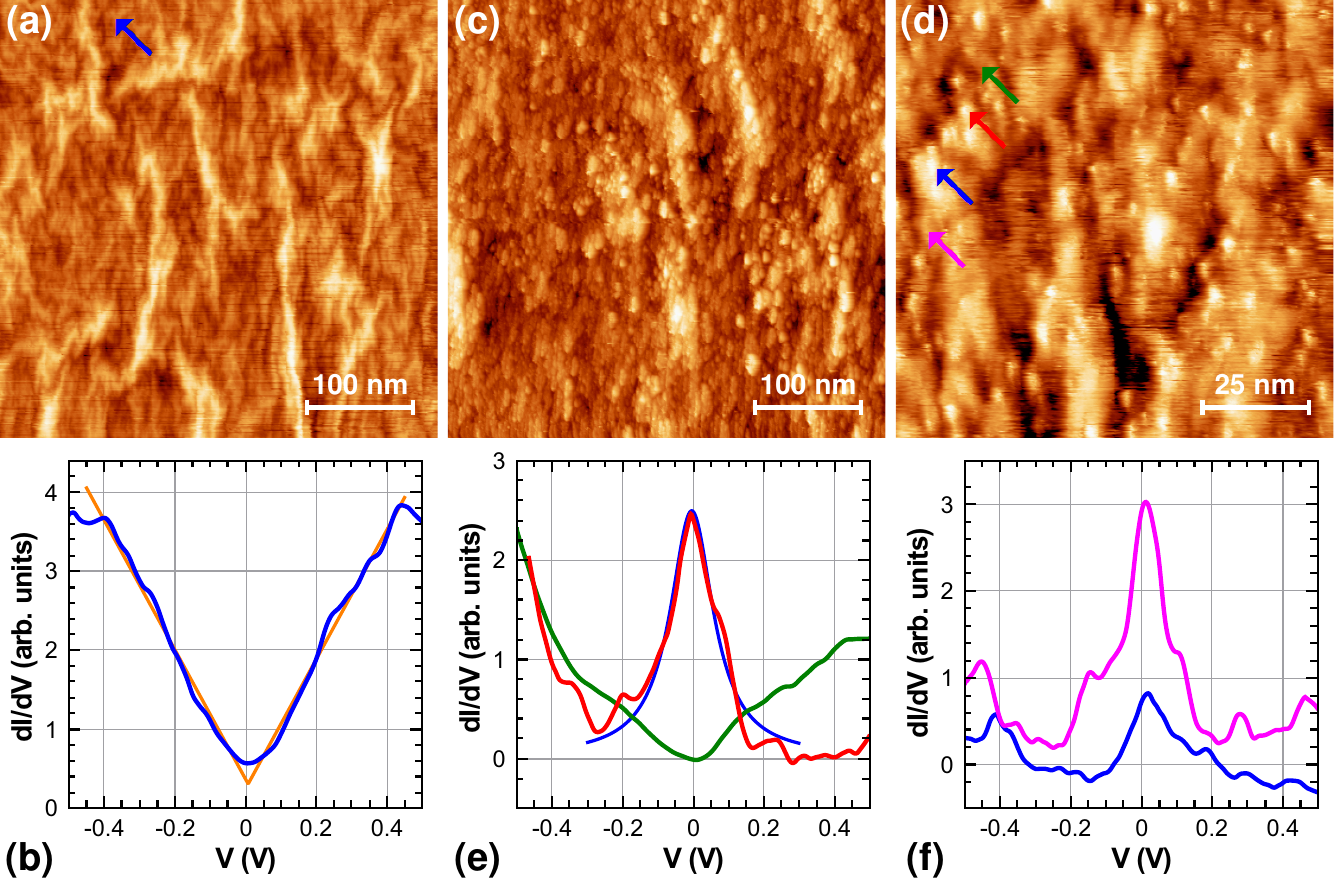}
\caption{(color online) {\it Graphene on SiO$_2$}: (a) STM image prior to ion bombardment with arrow marking the position of the $dI/dV$ curve in (b), $I=0.1\rm\,nA$, $V=-900\rm\,mV$; (b) $dI/dV$ curve (blue curve) recorded at the arrow in (a) with linear guides to the eye (orange lines), $I_{\rm stab}= 0.1$ nA, $V_{\rm stab}=-900$ mV, $V_{\rm mod}=20$ mV; (c) STM image after Ar$^+$ bombardment at $E_{\rm kin}=50$ eV with fluence: $8\cdot 10^{-3}$ ions/nm$^2$, $I=0.1\rm\,nA$, $V=-800\rm\,mV$; (d) STM image after Ar$^+$ fluence: $1.0$ ions/nm$^2$, $E_{\rm kin}=50$ eV, $I=0.04\rm\,nA$, $V=-120\rm\,mV$; arrows mark positions of $dI/dV$ curves in (e) and (f); (e), (f) $dI/dV$ curves recorded at the points marked by arrows with the same color in (d), blue curve in (e) is a Lorentzian fit to the red curve used to determine peak energy and width,  $I_{\rm stab}= 0.04$ nA, $V_{\rm stab}=-120$ mV, $V_{\rm mod}=20$ mV.}
\label{Fig2}
\end{figure*}
%
%, where the uncertainty is caused by the unprecise determination of the image sizes due  to creep and thermal drift.
The defects of our STM study, in addition, mostly exhibit a peak around $E_D$ in $dI/dV$ curves. Figure \ref{Fig1}(b) shows $dI/dV$ curves obtained on different positions of an individual defect in comparison to a curve obtained on the flat HOPG surface. The defect-related peak is apparent. It slightly changes in position and more strongly in intensity across the defect. The peak position fluctuates from defect to defect with an average peak position of $-1$ mV  and a rms fluctuation of 30 mV. The average full width at half maximum (FWHM) of the peak obtained from Lorentzian fits is 110 mV with a rms fluctuation of 20 mV. Figure \ref{Fig1}(c) and (d) show $dI/dV$ images at low (-10 mV) and high (-160 mV) bias revealing a contrast inversion at the defect sites, which shows that all defects exhibit an increased LDOS around $E_{\rm F}$. The $dI/dV$ curve away from the defects
(blue curve in Fig. \ref{Fig1}(b)) shows a minimum close to $E_{\rm F}$ indicating minimal doping by the ion bombardment, i.e. $E_{\rm D}\simeq E_{\rm F}$ with the Fermi level $E_{\rm F}$.
%Thus, the defects are very similar to the ones observed on graphene (Fig. {\ref{Fig1}}(a)-(b)). The fluctuation in peak position is slightly reduced, which might be related either to the rippling or to the more inhomogeneous electrostatic environment of the graphene sample.
\\
The peaks at $E_{\rm D}$ are additional evidence that the defects are single vacancies, i.e. tight-binding calculations and previous STM results allow to identify single vacancies by a peak at $E_{\rm D}$ \cite{Pereira,Brihuega}. Thereby, the calculated vacancy induced peak at $E_{\rm D}$ remains largely unchanged up to 0.5 vacancies/nm$^2$ and gets  hybridized only at a density above 2 vacancies/nm$^2$, which is more than three orders of magnitude larger than in the experiment of Fig. \ref{Fig1} \cite{Ferro}. Different reconstructions of divacancies exhibit one or several peaks according to DFT, which are shifted by $0.2-0.8$ eV away from $E_{\rm D}$ \cite{Lherbier,Ugeda}. STM results which claim to have identified a divacancy produced by 140 eV Ar$^+$ bombardment find a multiple peak in the LDOS about 0.2 eV above $E_{\rm D}$ \cite{Ugeda}. Thus, the only reasonable explanation for the peak at $E_D$ is a single vacancy produced by ion bombardment.\\
%SRIM simulations  of a single layer of amorphous carbon on SiO$_2$ indeed exhibit a sputter yield of only $8 \cdot 10^{-5}$ atoms/ion and a carbon displacement yield of $0.33$ vacancies/ion.
\\
Using this sample, we tested the development of the defects under air exposure. Figure \ref{Fig1}(e) and (f) show a STM image and $dI/dV$ curves, respectively, after taking the sample out of the UHV and exposing it to air for 10 min. The STM image features a double tip, i.e all defects are accompanied by a fainter ghost image at the upper left. However, this does not influence the spectroscopic characterization, since the two features of the double tip are spatially well separated, i.e. the double tip only mixes contributions from clean graphene into the spectra on the defects. Figure \ref{Fig1}(g) and (h) show the same data after 3 hours of air exposure. The average peak value and average peak width are -10 mV (-20 mV) and 130 mV (140 mV) after 10 min (3 hours) with the same rms fluctuations as prior to the air exposure. Thus, the peak gets slightly broadened by air exposure, but remains close to $E_{\rm F}$ indicating the persistence of paramagnetic properties. The fact that the peak gets slightly broader on average is probably caused by
the onset of an interaction with adsorbates from air. This is in line with the observation that extensive air exposure also broadens the ESR line (see below). A speculation, which type of adsorbate is responsible for the slight broadening is beyond the scope of this manuscript. Importantly, we find that the spectral change of the defect is minor such that defects characterized by peaks in the local density of states close to $E_D$, being most likely reconstructed single vacancies, can be probed by ESR even after short term transfer through air.\\
\begin{figure*}  % B_Mod = amplitude
\includegraphics[width=17.8cm]{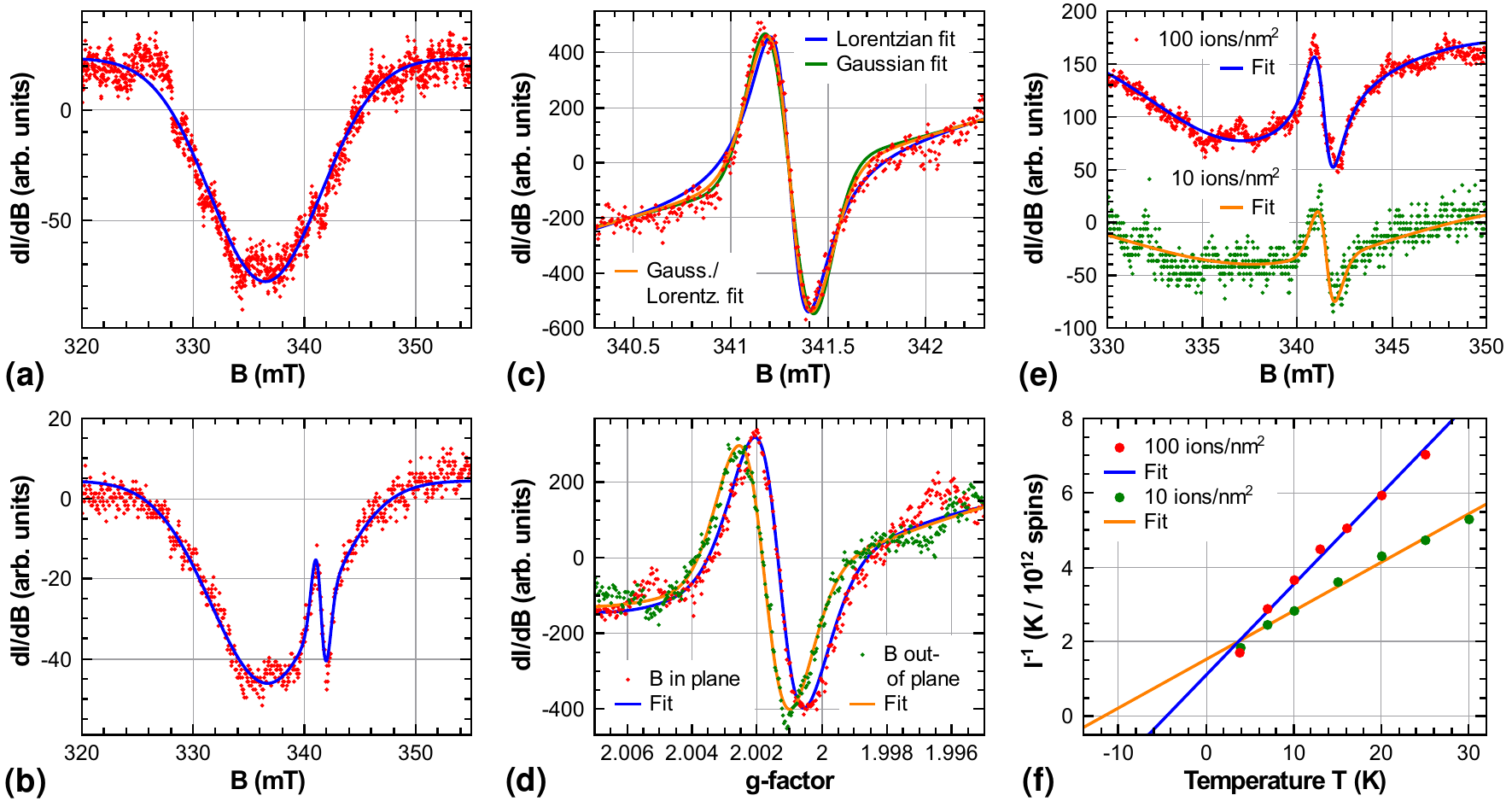}
\caption{(color online)
{\it ESR spectra after Ar$^+$ bombardment on graphene, $E_{\rm kin}=50$ eV}: (a) sample without Ar$^+$ bombardment, Gaussian fit curve (blue line) is added, which is used as an offset for the fit curves in (b)-(e), $f=9.5720\rm\,GHz$,
  $B_{\rm mod}=0.8\rm\,mT$; (b) sample at Ar$^+$ fluence of $10$ ions/nm$^2$, $f=9.5726\rm\,GHz$,
  $B_{\rm mod}=0.5\rm\,mT$; (c) sample at Ar$^+$ fluence of $10$ ions/nm$^2$, $f=9.5619\rm\,GHz$,
  $B_{\rm mod}=0.05\rm\,mT$ with fit curves as marked (see text); (d) sample at Ar$^+$ fluence of $10$ ions/nm$^2$ with $B$-field applied perpendicular and parallel to the sample plane as indicated and mixed fit curves, $f=9.5611\rm\,GHz$ ($f=9.5622\rm\,GHz$) for in-plane (out-of-plane) field,
  $B_{\rm mod}=0.1\rm\,mT$, due to the different measurement frequencies for the two curves, the $x$-axis displays the corresponding $g$ values; (e) sample at different Ar$^+$ fluence as indicated after optimizing the sample contacts by heating to 200$^\circ$ C in Ar atmosphere and additional air exposure of about 60 min, Lorentzian fit curves (lines) are added, $10$ ions/nm$^2$: $f=9.5716\rm\,GHz$, $B_{\rm mod}=0.5\rm\,mT$;  $100$ ions/nm$^2$: $f=9.5760\rm\,GHz$,
  $B_{\rm mod}=0.8\rm\,mT$; (f) temperature dependent inverse peak area of the ESR curves calibrated by a ruby standard (see text) in comparison with fit curves $a\cdot (T-\theta_{\rm cw})$ revealing $\theta_{\rm cw}=-12$ K and $\theta_{\rm cw}=-5$ K, respectively.}
\label{Fig3}
\end{figure*}
\subsection{STM results on graphene}
Figure \ref{Fig2}(a) shows a STM image of the graphene sample prior to ion bombardment. The large folds with heights up to 2 nm probably originate from the transfer process. They are accompanied by ripples on the scale of 10 nm as similarly observed on graphene flakes on SiO$_2$ prepared by the scotch tape method \cite{Stolyarova}. The corrugation exhibits a rms roughness of 390 pm. The $dI/dV$ spectroscopy [Fig. \ref{Fig2}(b)] shows a rather linear increase on the hole and on the electron side with interception around $V=0\pm 10$ mV revealing that the sample is barely doped. After ion bombardment [Fig. \ref{Fig2}(c),(d)], the additional white bumps indicate the ion induced defects.
Such bumps have been observed previously on graphene on SiC and SiO$_2$ after ion bombardment at higher energy \cite{Ugeda,Tapazto}. Counting the defects on the graphene sample is not very reliable due to the additional contrast of the rippling and the folds. At low fluence [Fig. \ref{Fig2}(c)], we estimate a
yield of about $Y\simeq 0.1$ defects/ion very similar to the one obtained for HOPG.  Based on the same arguments as for HOPG, we conclude that the majority of the defects are single vacancies. Notice that MD simulations exclusively find single vacancies after Ar$^+$ bombardment at 50 eV of monolayer graphene \cite{Lehti} and again that the TEM data imply that the single vacancies are immobile \cite{Kaiser}. Indeed, the defects exhibit a peak in $dI/dV$ spectroscopy exclusively around $E_{\rm F}$ up to ion fluences of 1/(nm$^2$) as expected for single vacancies [Fig. \ref{Fig2}(e),(f)]. At fluences higher than 1/nm$^2$, it gets difficult to spot areas without $dI/dV$ peaks. The averaged peak position (16 spectra) is $-5$ mV with a peak to peak variation of 50 mV$_{\rm rms}$. The FWHM of the peak is $110$ mV with a rms fluctuation of 50 mV.  Thus, the defects are very similar to the ones observed on HOPG (Fig. {\ref{Fig1}}). The rms fluctuation in peak position is slightly stronger on graphene, which might be related either to the rippling or to the more inhomogeneous electrostatic environment of the graphene sample. The Dirac point corresponding to the minimum of the $dI/dV$ curves recorded away from the defects barely shifts again indicating negligible doping by ion bombardment.\\

\subsection{ESR results on graphene}
In this section, the key results obtained on the samples characterized by STM are described. Since DFT also predicts a magnetic moment of $1 \mu_{\rm B}$ for a single vacancy after Jahn-Teller distortion \cite{Lehtinen} accompanied by the LDOS peak at $E_{\rm D}=E_{\rm F}$  \cite{Ferro}, the magnetic moment should be observable in ESR experiments. Such measurements are displayed in Fig. \ref{Fig3} as recorded at $f=9.56-9.58$ GHz in the derivative mode. Figure \ref{Fig3}(a) probes a graphene sample prior to ion bombardment. It exhibits only a broad dip, which is also present without graphene, and, thus, a resonance feature of the sample holder or the substrate. This feature is not considered for further analysis. Spectra at low fluence did not show any additional signatures. For an ion fluence of 10/nm$^2$ [Fig. \ref{Fig3}(b)], we find an additional narrow ESR line corresponding to $g\simeq 2.002$ as expected for graphene vacancies \cite{Lee}.  The apparent line width can be determined from a fit, which is optimal, if a mixture of a Gaussian and a Lorentzian of similar strength are used. Fit curves of pure Gaussian and pure Lorentzian type are shown in comparison to the mixed fit in Fig. \ref{Fig3}(c). While the Lorentzian fit deviates at low magnetic field from the experimental data, the Gaussian fit deviates at high magnetic field, which can be both compensated by the mixture (Gauss./Lorentz. fit). This indicates that both, broadening due to inhomogeneities of the local $g$-factor and anisotropic dipole-dipole interaction (Gaussian) as well as exchange narrowing (Lorentzian) contribute to the line width. The latter one means that the isotropic exchange between the electrons averages the local fields from neighboring spins, which results in a narrowing of the line width, which then partly exhibits the Lorentzian shape due to the finite lifetime (see also below).
The apparent width of the peak, i.e. the distance between the minimum and the maximum of the fitted derivative curve, is 0.18 mT. Considering the influence of $B_{\rm mod}$, this translates to an intrinsic width of 0.17 mT.\\
Figure \ref{Fig3}(d) shows ESR spectra of the same sample with the $B$-field in-plane and out-of plane. A small shift of the ESR line is observed. The fit curves to the data, which are also displayed, reveal a $g$-factor of 2.0013 and 2.0018 for in-plane and out-of-plane direction of the field, respectively, i.e. an anisotropy of 0.02 \%.\\
% or a quality factor of the resonance of about 2000. This quality factor is more than an order of magnitude larger than after 225 MeV proton bombardment of HOPG at an ion fluence of $2\cdot 10^{21}$ m$^{-2}$ \cite{Lee}. It is also better than for most of the non-bombarded graphene or nano-graphite samples studied previously \cite{Ciric,Ciric2,Ciric3,Kaustelkis, Smith, Shibayama, Lijewski, Andersson, Kempinski, Joly, Joly2, Rao1, Rao2, Rao3, Rao4, Tommasini, Fabian, Ciric4, Akbari}. Only the monolayer platelets on SiO$_2$ revealed even higher quality factors of 6000-20.000 \cite{Yablokov1,Yablokov2}.\\
The sample of Fig. \ref{Fig3}(c)$-$(d) was exposed to an ion fluence of $10$/nm$^2$ and subsequently to air for about 30 min prior to the ESR measurements.
The ESR curves in Fig. \ref{Fig3}(e) are recorded for the two ion fluences marked, but after an additional heating to 200$^\circ$ C in Ar atmosphere for 20 min for both samples and an additional exposure to air for 60 min for the sample with fluence of $10$/nm$^2$. This additional treatment is required for optimal electrical contacting. The ESR peak barely shifts but gets significantly broader. The same behaviour was observed for the sample shown in Fig. \ref{Fig3} (b) after additional air exposure of about 60 min without heating. In these three cases, the intrinsic width is about 0.7 mT (we checked that reducing $B_{\rm mod}$ did not change the peak width) indicating that the width depends more on preparational details than on the ion fluence. A detailed investigation on the origin of the linewidth broadening is beyond the scope of this study. It might be that some of the spins are quenched by interaction with adsorbates or by recombination with nearby vacancies leading to a reduced exchange narrowing (see below) and/or to an increased inhomogeneity of the local g-factor. Importantly, a line width as small as 0.17 mT appears after ion bombardment and fast transfer through air.\\

The number of spins $N_{\rm G}$ contributing to the ESR signal of graphene can be estimated by comparison with a ruby standard with calibrated spin number $N_{\rm R}$ \cite{Chang}. One has to consider the different strength of the microwave magnetic field $B^{\rm mw}$ at the position of the ruby crystal $B^{\rm mw}_{\rm R}$ and at the position of graphene $B^{\rm mw}_{\rm G}$  and the different spin transition probabilities $P_{\rm R}$ and $P_{\rm G}$. The data for ruby are tabulated at room temperature \cite{Chang}, while the graphene samples exhibit ESR signals only at low temperature, such that the temperature difference must be considered, too. We calibrated the $B^{\rm mw}$ field ratio $B^{\rm mw}_{\rm G}/B^{\rm mw}_{\rm R}$ by a primary measurement at 300 K, where one ruby crystal was placed at the
later graphene position and the other at the later ruby position, which exhibits $T_{\rm R}=300$ K even if the flow cryostat is operated. We calculate the areas $A_{\rm G}$ and $A_{\rm R}$ under the ESR absorption peaks of graphene and ruby, respectively, as usual, from the square of the peak to peak linewidth multiplied by the peak to peak amplitude of the derivative signal. Thereby, for graphene we assumed a mixture of equal contributions from a Gaussian and a Lorentzian as implied by the fit. The last step causes an error of about 50 \% due to the uncertainty in the relative strength of the two contributions.
$N_G$ at graphene temperature $T_{\rm G}$ is then deduced from:
\begin{equation*}
\frac{A_{\rm G}}{A_{\rm R}}= \left(\frac{B^{\rm mw}_{\rm G}}{B^{\rm mw}_{\rm R}}\right)^2\cdot\frac{g_{\rm G}(2S_{\rm R}+1)}{g_{\rm R}(2S_{\rm G}+1)}\cdot\frac{P_{\rm G}}{P_{\rm R}}\cdot \frac{N_{\rm G}}{T_{\rm G}-\theta_{\rm cw}} \cdot \frac{T_{\rm R}}{N_{\rm R}}
\end{equation*}\\
where $S_{\rm G}=\frac{1}{2}$ for graphene and $S_{\rm R}=\frac{3}{2}$ for the Cr$^{3+}$ ions in the ruby crystal. The missing $P_{\rm G}$ is calculated using the matrix element of a spin 1/2 system with negligible anisotropy \cite{Abragam}. The Curie-Weiss temperature $\theta_{\rm cw}$ takes into account possible magnetic correlations between graphene vacancies according to the Curie-Weiss law.\\
Figure \ref{Fig3}(f) shows the plot of the inverse intensity $I^{-1}:=(N_{\rm G}/(T_{\rm G}-\theta_{\rm cw}))^{-1}$ as a function of $T_{\rm G}$ for an Ar$^+$ fluence of 10 ions/nm$^2$ and 100 ions/nm$^2$ as marked. The linear fits allow extraction of $N_{\rm G}$ and $\theta_{\rm cw}$. We get $N_{\rm G}=(7.7\pm 4.8)\cdot10^{12}$ and $\theta_{\rm cw}=-11.7\pm 1.4$ K for a fluence of 10 ions/nm$^2$ and $N_{\rm G}=(4.1\pm 2.5)\cdot10^{12}$ and $\theta_{\rm cw}=-4.5\pm 0.7$ K for a fluence of 100 ions/nm$^2$. Since $\theta_{\rm cw}$ is negative, the dominating correlations are obviously antiferromagnetic and not ferromagnetic. We do not observe any ordering transition down to $T_{\rm G}=4$ K, which would exhibit a  broadening and possibly a shift of the ESR line.\\
From the Curie-Weiss temperature $\theta_{\rm cw}$ we can also estimate the exchange integral $J=(3k_{\rm B}\theta_{\rm cw})/(2zS_{\rm G}(S_{\rm G}+1))$ \cite{Kittel}. Assuming the number of nearest neighbours on a honeycomb lattice $z=3$ we obtain $J= 0.7$ meV and 0.3 meV for the two samples, respectively. \\

The spin densities are determined from $N_{\rm G}$ by dividing through the area of the graphene flakes measured with an optical microscope. They are $2\pm 1$/nm$^2$ and $0.5\pm 0.2$/nm$^2$, i.e. the average distances $d_{\rm s}$ amount to $d_{\rm s}=0.7$ nm and $d_{\rm s}=1.4$ nm, respectively.\\
The anisotropic dipole-dipole interaction contributes to the broadening of the ESR signal and yields Gaussian line shapes \cite{Poole}. The widths $\Delta B_{\rm dd}$ from dipole-dipole interaction  would be approximately $0.2-0.03$ T, i.e.  $2-3$ orders of magnitude larger than in the experiments. Since, we observe a mixture of Gaussian and Lorentzian shapes, the corresponding width has to be corrected by the narrowing effect of the isotropic exchange interaction $J$:
$\Delta B \approx \Delta B^2_{\rm dd}/(1.5\frac{\mu_0J}{g\mu_{\rm B}})$ \cite{Poole}.
With the observed values for the line width $\Delta B$ and the exchange integral $J$, the dipole-dipole line width can be further used for an independent estimation of the average distances between the spins, since $\Delta B_{\rm dd}$ is straightforwardly related to $d_{\rm s}$ [67]. The resulting values are $d_{\rm s}= 0.64$ nm for the first sample ($N_{\rm G}=7.7\cdot 10^{12}$, $J=0.7$ meV ) and $d_{\rm s}=0.75$ nm for the second one ($N_{\rm G}=4.1\cdot 10^{12}$, $J=0.3$ meV). These results are in fact self-consistent: the smaller distance between the spins yields a stronger exchange coupling $J$. They also agree with the above estimate from the spin density and confirm independently that the mean distance between the spins participating in the resonance is $d_{\rm s} \simeq 1$ nm.\\
%The temperature dependence of the ESR peak area is deduced from fits to the data as displayed, e.g., in Fig. \ref{Fig5}(c). While the peak width changes by less than 25 \% from 4 K to 20 K, the peak area changes by more than a factor of two. The ESR peak area $A$ is displayed as a function of temperature $T$ in Fig. \ref{Fig5}(g) together with a fit curve using the Curie-Weiss law $A =1/(a\cdot (T-\Theta_{\rm p}))$ with fit parameters $a$ and $\Theta_{\rm p}$. The fit reveals $\Theta_{\rm p}=4\pm 1$ K indicating a slight tendency for antiferromagnetism in accordance with \cite{Yablokov1,Yablokov2}, but more importantly excludes ferromagnetism at higher temperatures after an ion fluence of 10/nm$^2$.
The spin yield observed by ESR at an ion fluence of 10/nm$^2$, $Y\simeq 0.2$ spins/ion, is similar to the vacancy yield observed in STM at lower fluence ($Y\simeq 0.1$). This could imply that each vacancy contributes a spin 1/2 to the signal. However, the strong reduction of the spin yield at a fluence of 100/nm$^2$ ($Y\simeq 0.01$ spins/ion) points to a significant self-healing of the graphene by nearby ion impacts, which might also be partly present at the fluence of 10/nm$^2$. Thus, spin 1/2 can only act at a lower bound for the spins per vacancy.\\
The self-healing might be explained as follows.
The small average distance between defects of $0.5-1$ nm leads to the formation of divacancies lacking a magnetic moment, since the energy gain in divacancy formation is more than 3 eV per vacancy according to DFT \cite{Krashen}. This implies even a barrierless formation at close enough distance. Moreover, DFT finds that the so-called 585 reconstruction of a divacancy exhibits a spin $S=0$ and a LDOS peak at $-0.3$ eV \cite{Lherbier}. Indeed, we occasionally observe $dI/dV$ peaks around -0.2 eV on the graphene surface after Ar$^+$  bombardment with fluence 10/nm$^2$.\\
Notice finally, that DFT predicts antiferromagnetic order between vacancies on different sublattices and ferromagnetic order between vacancies on the same sublattice \cite{Yazyev}. The Hubbard model implies a significantly stronger antiferromagnetic coupling by superexchange between different sublattices than the ferromagnetic coupling by direct exchange on the same sublattice \cite{Li}. The superexchange for small distances ($< 10$ lattice sites) is also much larger than the exchange via RKKY interaction reaching about $J=150$ meV at a distance of 5 lattice sites \cite{Li}. Thus, the experimentally found preferential antiferromagnetic correlations between vacancies are in qualitative agreement with the theoretical predictions. However, the calculated interaction would lead to much higher Curie-Weiss temperatures, at least, if the compensating effects of the ferromagnetic interactions and the disorder are not taken into account. This calls for more detailed investigations also from the theoretical side.\\
\begin{figure}
\includegraphics[width=7.0cm]{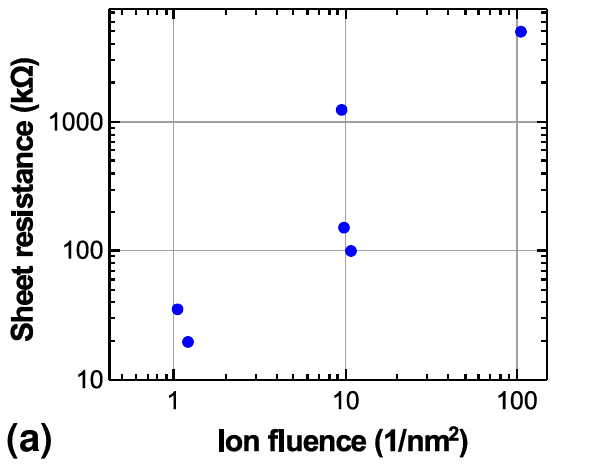}
\caption{(color online) Sheet resistance of monolayer graphene on SiO$_2$ as a function of ion fluence measured at 300 K}
\label{Fig4}
\end{figure}
\subsection{Transport and EDSR results on graphene}
Finally, we comment on our attempt to measure changes of the sample conductivity during $\mu$-wave exposure in changing $B$-field aiming for EDSR. While the sheet resistance $\rho$ of graphene increases with increasing ion fluence $F$ up to 3.5 G$\Omega$ (slope $\rho\propto F^{1.1}$) [\ref{Fig4}] (see also \cite{Fuhrer}), we do not detect a relative current change by applying the same oscillating $B$ field as in the ESR experiments of Fig. \ref{Fig3}(c), down to $\Delta I_{\rm ac}/I_{\rm dc} = 10^{-5}$ with $I_{\rm dc}$ being the applied current and $\Delta I_{\rm ac}$ the amplitude of the current change. This implies that the spin orientation of the single vacancies is not relevant for transport.\\

\section{Conclusion}
In conclusion, we have presented a combined study of electron spin resonance and scanning tunneling spectroscopy on the same graphene samples after low energy ion bombardment.
We detect the LDOS peak close to the Dirac point indicative of paramagnetic single vacancies, which persists air exposure up to 3 hours. Electron spin resonance data exhibit a resonance line close to $g=2.002$ with an anisotropy of 0.02 \% only. A Curie-Weiss-type temperature dependence with a Curie-Weiss temperature of about -10 K proofs the existence of preferential antiferromagnetic correlations at defect densities of $1-2$/nm$^2$. This excludes ferromagnetism even at low temperature. We regard these results as an important step towards a more controlled investigation of defect induced magnetism in graphene. Moreover, these samples might be a good benchmark for open questions concerning ESR-STM measurements \cite{Yishai}.

\section{Acknowledgment}
We gratefully acknowledge helpful discussions with M. Liebmann and P. Nemes-Incze, technical assistance of M. Grob, N. Freitag, and K. Fl\"ohr, and
financial support by the Graphene Flagship (contract no. NECT-ICT-604391) as well as by the German science foundation via FOR 912 "Coherence and relaxation properties of electron spins".

\end{document}